\documentstyle [twoside,12pt]{article}
\newcommand{\nc}{\newcommand}
\nc{\la}{\lambda} \nc{\La}{\Lambda}  \nc{\al}{\alpha}
\nc{\te}{\theta}  \nc{\be}{\beta}	\nc{\ka}{\kappa}
\nc{\ga}{\gamma}  \nc{\Ga}{\Gamma}
\nc{\de}{\delta}  \nc{\De}{\Delta}
\nc{\si}{\sigma}  \nc{\Si}{\Sigma}
\nc{\om}{\omega}  \nc{\Om}{\Omega}
\nc{\nf}{\infty}   \nc{\nl}{\newline}
\nc{\ra}{\longrightarrow}
\nc{\beq}{\begin{equation}}
\nc{\eeq}{\end{equation}}
\nc{\beqa}{\begin{eqnarray}}  \nc{\dst}{\displaystyle}
\nc{\eeqa}{\end{eqnarray}} \nc{\nnb}{\nonumber}
\title{\bf Lorentz and  CPT violations in QED : \\ a short comment
on recent controversies.}\author{Guy Bonneau\thanks {\noindent
Laboratoire de Physique Th\'eorique et des Hautes Energies,
 Unit\'e associ\'ee au CNRS UMR 7589, Universit\'e Paris 7,
 2 Place Jussieu, 75251 Paris Cedex 05. Email:
bonneau@lpthe.jussieu.fr}}
\begin{document}
\date{}
\maketitle
\begin{abstract}
\noindent We rediscuss the recent controversy on a possible
Chern-Simons like term generated through radiative corrections in
QED with a CPT  violating term. We emphasize the fact that any
absence of an {\sl a priori} divergence should be explained by some
symmetry or some non-renormalisation theorem : otherwise, no
prediction can be made on the corresponding quantity.
\end{abstract}

\vspace{5cm}



\vfill {\bf PAR/LPTHE/01-36}\hfill  September 2001
\newpage

In the past years, the interesting issue of a possible spontaneous
breaking of Lorentz invariance at low energy has been considered :
this issue also led to CPT breaking
\cite{{CFJ90},{ColKos},{ColGlas}}. In particular, the general
Lorentz-violating extension of the minimal
$SU(3)\times SU(2)\times U(1)$ standard model has been discussed :
as many breaking terms are allowed, people look for possible
constraints coming from experimental results as well as from
renormalisability requirements and anomaly cancellation.

In that respect, there arose a controversy on a possible
Chern-Simons like term generated through radiative corrections
[2-16]. This phenomenon was
studied in QED, an abelian gauge theory, as a part of the standard
model. The aim of this note is to clarify the discrepancies and insit
upon the need of an explanation when some one-loop divergence does
not appear.

 The Lagrangian density is~:

\beqa\label{QED1} {\cal L} & = & {\cal L}_0 + {\cal L}_1 + {\cal
L}_2 \\ a)\ {\cal L}_0 & = & \bar{\psi}(i\not{\partial} - m -
e\not{A})\psi -
\frac{1}{4}F_{\mu\nu}^2 -\frac{1}{2\al}(\partial A)^2
+\frac{1}{2}\la^2 A_{\mu}^2 \nnb\\
\quad & \quad & {\rm where}\ \al\ {\rm is\ the\ gauge\ parameter\
and}\ \la\ {\rm an\ infra-red\ regulator\ photon\ mass\,,} \nnb \\
b)\  {\cal L}_1(x) & = & -
b^{\mu}\bar{\psi}(x)\ga_{\mu}\ga^5\psi(x)\,,\quad\quad {\rm where}\
b^{\mu}\ {\rm is\ a\ fixed\ vector,}\nnb\\  c)\ {\cal L}_2(x) & = &
\frac{1}{2}c^{\mu}\epsilon_{\mu\nu\rho\si}
F^{\nu\rho}(x)A^{\si}(x)\,,\quad\quad {\rm where}\  c^{\mu}\ {\rm
is\ a\ fixed\ vector.} \nnb \eeqa Other breakings could be
considered (see a discussion in the first paper of \cite{ColKos}),
but we simplify and require charge conjugation invariance, which
selects
${\cal L}_1(x)$ and ${\cal L}_2(x)\,.$ Note for further reference
that experiments on the absence of birefringence of light in vacuum
put very restrictive limits on the value of $c^{\mu}\,,$ typically
for a timelike
$c^{\mu}\,,\ c^{\small 0}/m \le 10^{-38 }$ \cite{ColKos}.
\begin{itemize}
\item In order to avoid the difficulties resulting from the new
poles in the propagators, as in \cite{Bonneau01}, we take into
account the smallness of the breakings and include them into the
interaction Lagrangian density as super-renormalisable couplings
(see in \cite{ColKos} a discussion on the alternative choice
of a complete propagator). Moreover, we define
the photon and electron masses by the same normalisation conditions
as in ordinary Q.E.D.,
$e.g.$
$$\dst <\psi(p)\bar{\psi}(-p)>^{prop.}\mid_{\not{p} = m\,,\;b = c
=0}\quad =\ 0\,,\quad\cdots$$ According to standard results in
renormalisation theory, these breakings add new terms in the
primitively divergent proper Green functions. By power counting, these are
$$\Ga_{\mu\nu} (p,-p)\,,\
\Sigma (p,-p)\,,\ \Ga^{\rho} (p,q,-(p+q))\ {\rm and}\
\Ga_{\mu\nu\rho\si} (p_1,\,p_2,\,p_3,\,-(p_1+p_2+p_3))\,,$$
 respectively the photon and electron 2-points proper Green
functions, the photon-electron proper vertex function and the photon
4-point proper Green function.  The corresponding overall  divergences (sub-divergences being properly subtracted) are
polynomial in the momenta and masses
\footnote{\
 C invariance has been used. The Ward identity (\ref{WI4}) will
relate some of these parameters : $a_2 = a_3 = a_4 = 0\,,\ a_{11} = e a_7\,,\ a_{12} = 0. $}:

\beqa\label{div1}\Ga_{\mu\nu}(p,-p)\mid_{div} & = & a_1[g_{\mu\nu}
p^2 - p_{\mu}p_{\nu}] + a_2 p_{\mu}p_{\nu} + [a_3 m^2 + a_4
\la^2]g_{\mu\nu} + [a_5 b^{\rho} + a_6
c^{\rho}]\epsilon_{\mu\nu\rho\si}p^{\si}\,,\nnb\\
\Sigma(p,-p)\mid_{div} & = & a_7 \not{p} + a_8 m + [a_9 b^{\rho} +
a_{10} c^{\rho}]\ga_{\rho}\ga^5 \,,\nnb\\
\Ga^{\rho}(p,q,-(p+q))\mid_{div} & = & a_{11} \ga^{\rho}  \,.\nnb\\
\Ga_{\mu\nu\rho\si}(p_i)\mid_{div} & = & a_{12}
[g_{\mu\nu}g_{\rho\si} + g_{\mu\rho}g_{\nu\si} +
g_{\mu\si}g_{\rho\nu}]\,.\eeqa
 All  parameters
$a_i[e, \frac{b^2}{m^2}, \frac{c^2}{m^2},\ \frac{b\cdot c}{m^2}]$ (positions and residues of the poles in propagators, couplings
at zero momenta,..) - but for the unphysical, non renormalised ones
(as the longitudinal photon propagator (gauge parameter $\al$) and
photon regulator mass
$\la^2$ for unbroken QED) - require normalisation conditions, a
point which has often been missed since the successes of {\bf
minimal} dimensionnal regularisation scheme
\cite{GB1990} but is stressed in the recent reviews
\cite{{Victoria},{Chen2001}} and in a computation on QED at finite temperature \cite{Cervi}.
 In particular we shall require 2 new normalisation conditions to
fix the breaking parameters
$b^{\mu}$ and
$c^{\nu}\,:$
\beqa\label{nor1} b^{\mu} &  = & - \frac{i}{4}
Tr[\ga^{\mu}\ga^5<\psi(p)\bar{\psi}(-p)>^{prop.}]\mid_{p=0}\,,\nnb \\
  c^{\mu} & = &
\frac{1}{12}\epsilon^{\mu\nu\rho\si}\frac{\partial}{\partial p^{\si}}
<A_{\nu}(p)A_{\rho}(-p)>^{prop.}\mid_{p=0}\,.\eeqa

\noindent Note that, contrary to  ${\cal L}_1(x)\,,$ the ${\cal
L}_2(x)$ term also breaks the local gauge invariance of the
Lagrangian density, but we emphasize the fact that - except for the unphysical part $\dst\int [ -\frac{1}{2\al}(\partial A)^2
+ \frac{1}{2}\la^2 A_{\mu}^2 ] $ - the action
$\Gamma = \int \cal{L}$ is invariant under local gauge
transformations.

Then a Ward identity may be written  :
\beqa\label{WI4}
\lefteqn{\int d^4 x
\left\{\frac{1}{e}\partial_{\mu}\La(x)\frac{\de\Ga}{\de A_{\mu}(x)}+
i\La(x)[\bar{\psi}(x)\frac{\stackrel{\rightarrow}{\de}\Ga}
{\de\bar{\psi}(x)} -
\frac{\Ga\stackrel{\leftarrow}{\de}}{\de{\psi}(x)}\psi(x)]\right\} =}
\nnb
\\  & =& \int d^4 x \left\{-\frac{1}{e\al}
\partial_{\mu}A^{\mu}(x)\Box\La(x)  +
\frac{\la^2}{e}A^{\mu}(x)\partial_{\mu}\La(x) +
\frac{1}{2e}\epsilon_{\al\be\de\mu}c^{\al}F^{\be\de}(x)
\partial^{\mu}\La(x)\right\}
\nnb
\\ & \Rightarrow & W_x\,\Ga\, \equiv \partial_{\mu}\frac{\de\Ga}{\de
A_{\mu}(x)} - ie[\bar{\psi}(x)\frac{\stackrel{\rightarrow}{\de}\Ga}
{\de\bar{\psi}(x)} -
\frac{\Ga\stackrel{\leftarrow}{\de}}{\de{\psi}(x)}\psi(x)] =
\frac{1}{\al} [\Box +
\al\la^2]\partial_{\mu}A^{\mu}(x)\,.
\eeqa Note that this equation is exactly the same as the one for
ordinary QED.

As soon as we use a regularisation that respects the symmetries
(gauge, Lorentz covariance and charge conjugation invariance), the
perturbative proof of renormalisability reduces to the check that the
${\cal{O}}(\hbar)$ quantum corrections to the classical action $\Ga\
:
\Ga_1 = \Ga|_{class.} + \hbar\De\,,$ constrained by the Ward identity
(\ref{WI4})  may be reabsorbed into the classical action through
suitable renormalisations of the fields and parameters of the
theory. This has been proved in \cite{Bonneau01}.

There, some local sources have been introduced to define the
{\bf local} operators ${\cal L}_1(x)$ and ${\cal L}_2(x)\,.$
Although this is only a technical tool, it has been criticised
\footnote{\ For example in page 3 of
\cite{Chen2001} : {\it `` Bonneau introduced external source fields
for the axial vector current and the CS term, so the Ward identities
he derived actually impose gauge invariance on Lagrangian density
..."} This assertion is wrong as the Lagrangian density is not gauge
invariant (moreover it has been gauge-fixed..) but, as proven in our
analysis \cite{Bonneau01}, the breaking of local gauge invariance is
a soft one and may be seen as a complementary part in the gauge fixing.}
so we now discuss whether the quantization of
$\Gamma =
\int
\cal{L}\,,$ without a {\bf local} definition of the operators
${\cal L}_1(x)$ and ${\cal L}_2(x)\,,$ is possible.

\item Of course, in ordinary QED, the axial current, being
uncoupled, is absent from the Lagrangian density and so does not need
to be defined as a quantum operator ; no axial vertex being present,
{\it a fortiori} there is no axial anomaly and no triangle graph to
consider.

On the contrary, in CPT-broken QED (\ref{QED1}), new axial
insertions enter the game, but they are integrated ones $\dst\int
d^4x{\cal L}_1(x)$ and $\dst\int d^4x{\cal L}_2(x)\,.$ Then, most of
the authors in that subject argue that introducing local sources for
the Lagrangian density breakings means adding supplementary
conditions on the theory.

\noindent I have two answers, or rather two objections :
\begin{itemize}

\item During my studies at University, I was not taught - nor
have I taught ! - how it is possible to {\bf define} the {\bf
space-time integration} of some local quantity if this quantity
itself is not defined at (nearly) every point of the space time
(notice that here the involved quantity is a product of quantum
fields at the same point of the space-time). On the contrary,
space-time integration,
$i.e.$ vanishing incoming momentum, sometimes introduces new
difficulties (IR divergences,...)\,! And, to define a quantum local
operator with the generating  functional approach, I know no other
way than the introduction of local sources, so it is what I did in
\cite{Bonneau01}.
\item Leaving aside this mathematical objection, suppose that one has only the
Ward identity (\ref{WI4}) at hand to constrain the possible
ultra-violet divergences (\ref{div1}). This is not sufficient to
prove that the breakings introduce no new infinities : in
particular, the Chern Simons term is of the right canonical
dimension and quantum numbers and
 satisfies  (\ref{WI4}) :
$$p^{\mu}\Ga_{\mu\nu}(p,-p) = 0 \quad \stackrel{ in\
particular}{\Rightarrow}  p^{\mu}<[\dst\int d^4x{\cal L}_1(x)]
A^{\mu}(p) A^{\nu}(-p)> = 0\,.$$ So, first, we have no explanation of
the fact that {\bf all one-loop calculations of the CS contribution
to the photon self-energy give a  finite result} ($a_5 = a_6 = 0$),
 second, being unconstrained, its finite part (renormalised value)
has to be fixed by a normalisation condition (a different situation
than a radiative correction such as the (g-2) or the Lamb-shift for
example). So, no prediction is possible and its value remains
arbitrary, which is rather unsatisfactory. 

I am really surprised
that in the twenty or so papers devoted to that subject, I could not
find one line of argument to explain this ``experimental "
one-loop\footnote{\ If this finiteness was an ``accidental" one, it would have no
reason to hold at higher-loop order\,!} finiteness, except in the
recent review by  P\'erez-Victoria
\cite{Victoria} where the finiteness of the CS term is related to
the one of the standard triangle graph in ordinary QED~: however, in
QED too, the finiteness of such a graph results from the gauge Ward
identity on the {\bf unintegrated axial vertex} 3-point function :
\beq\label{WIl} p_{\nu}<[\bar{\psi}\gamma^{\mu}\ga^5 \psi](-p-q)
A^{\nu}(p) A^{\rho}(q)> \ =\ q_{\rho}<[\bar{\psi}\gamma^{\mu}\ga^5
\psi](-p-q) A^{\nu}(p) A^{\rho}(q)>\ =\ 0\,.
\eeq

\end{itemize}

So, as long as no answer to these two objections has been given,
either you consider an ill-defined and no predictive theory of
broken-QED \footnote{\ Some authors rightly
conclude that the corresponding one-loop finite contribution is
ambiguous
\cite{{Jackiw},{Chung-Perez},{Victoria},{Chen2001}}.} or you agree to
consider an action in which every Lagrangian monomial is well-defined
as a quantum operator.

\vspace{0.5cm}

Then in \cite{Bonneau01}, I have proved that, being linear in the
quantum field, the variation of ${\cal L}_2(x)$ in a local gauge
transformation is soft : no essential difference occurs between
local  gauge invariance of the action and the ``softly" broken local
gauge  invariance of the Lagrangian density. As a consequence, the
theory (\ref{QED1}) is consistent (even with no ${\cal L}_2(x)$
term) and the CS term has been shown to be unrenormalised, {\bf to
all orders of perturbation theory}. So, its experimental ``vanishing"
offers no constraint on the other CPT breaking term
${\cal L}_1(x)\,.$

\item As a complement to the recent reviews
\cite{{Victoria},{Chen2001}}, let us now comment upon some points given
in the literature :

\begin{itemize}
\item Jackiw and  Kostelecky \cite{JK99} never introduce any
regulator. Then some of their relations are ``delicate ones" : see
for example for a divergent integral ( after equ.12), the
commutation of a derivation with respect to external momentum and
the integration \footnote{\, I cannot agree, and probably no
teacher can agree, with the comment on this work given in Chen's
review (\cite{Chen2001} p. 3): ``{\sl They ingeneously manipulated the
linear divergent term in the loop integration ..}". One should expect
that such ideas will disappear in published works.}. If the integral
in their equation (11) is computed with dimensional regularisation, a
result
$\dst\left[\frac{\theta}{\sin{\theta}} -1\right]$ is found, and not
simply
$\dst\frac{\theta}{\sin{\theta}}$ (with
$p^2 = 4 m^2\sin^2{\theta/2}\,).$

Moreover, in the absence of normalisation conditions or Ward
identities fixing some ambiguities, the difference of two equivalent
linearly divergent integrals gives an ambiguous logarithmic
divergent one. Even when one uses a symmetric integration that
suppresses the linearly (and eventually the logarithmicaly)
divergent part, the finite part remains ambiguous. The ``surface
term" that comes from a shift in the integration momentum in a
linearly divergent integral is a regulator dependent quantity :  if
one mimics the calculation in the appendix A5-2 of Jauch and
Rohrlich standard book
\cite{JR} with the dimensional scheme, one easily checks that no
"surface term" occurs after a shift of the integration momenta
\cite{DR2}). Recall that this possibility of shifting internal
momenta is needed to preserve gauge invariance in loop calculations
(see for example
\cite[subsect.17.9]{BD2}).

\item In a recent work \cite{Sitenko01}, the one-loop calculation of
the CS correction is done with the heat-kernel expansion and the
Schwinger proper-time method, leading to a new finite result,
claimed to be unambiguously determined. However,
\begin{itemize}\item  here again there is no explanation of the
absence of infinities in the result : then the finite part is {\sl a
priori} ambiguous
\footnote{\ Remember that in Fujikawa 's calculation of the axial
anomaly,  gauge invariance was implemented through the basis used to
compute the fermionic Jacobian : he chose eigenvectors of the
operator
$i\not{\partial} -e
\not{A}\,;$  another choice would allow the transfer of the axial
anomaly to some vector anomaly (see also the discussion on the
``minimal anomaly" in non-abelian gauge theory) \cite{Fujikawa}.},
\item other computations with the Schwinger proper-time method
exist \cite{proper-time} and give a different result, proving at
least that some ``ambiguity" remains,
\item some terms are lacking in
this calculation : in particular a logarithmically divergent
contribution to the CS term results from a thorough computation of
the quantity given in equation (21) of
\cite{Sitenko01} ( in the absence of any precise criteria to
substract infinite parts, this should not be a surprise).
\end{itemize}

\item It is difficult to see the difference often advocated  between
a first order (in
$b^{\mu})$ perturbative calculation and what is claimed to be a
``non-perturbative unambiguous value", but is, as a matter of fact,
obtained with exactly the same standard triangle integrals \cite{Jackiw}. Moreover, in
\cite{Chung-Perez} the computation is also done to all orders in the
breaking parameter $b^{\mu}$ and it is explicitely verified that
higher orders do not contribute to a possible correction to
$c^{\mu}\,.$

 However, in the first paper in \cite{ColKos}, Colladay and
Kosteleck{'}y gave a direct analysis of the {\bf complete} classical
fermion Green function  as defined by ${\cal L}_0 + {\cal L}_1\,.$ In
particular they check that the  anticommutator of two fermionic
fields vanishes for space-like separations, in agreement with
microcausality (at least for a time-like breaking $b_{\mu}\,).$ This
confirms our analysis on the correctness of a theory with no
classical CS term. Then, Adam and Klinkhamer show that the addition
of a ( radiatively generated) CS term ${\cal L}_2(x)\,$  with a
time-like $c^{\mu}$ breaks microcausality \cite{Adam}. As our
non-renormalisation theorem ensures that, if absent at the
classical level, the CS term will not appear in higher-loop order,
 microcausality will not be destroyed in higher-loop
order.

\end{itemize}
\end{itemize} Note also that the Lagrangian density
${\cal L}_0\ + {\cal L}_2$ would not lead to a coherent theory as an
(infinite) counterterm ${\cal L}_1$ appears at the one-loop order
\cite{Bonneau01}.

\vspace{1cm}

\noindent To summarize, we have proven that : \begin{itemize}
\item The local gauge invariance of the Lagrangian density  is  destroyed by a ${\cal L}_2$ term (plus of course by the usual gauge fixing term) : but, being bilinear in the gauge field, ${\cal L}_2(x)$ behaves as
a minor modification of the {\bf gauge fixing term} as
$\partial_{\nu}A^{\nu}$ remains a free field. As part of the
``gauge term", this ${\cal L}_2(x)$ is,  as usual, not renormalised :
so its all-order value is equal to its (arbitrarily chosen)
classical one.
\item A theory with
a vanishing tree level Chern-Simons like breaking term is consistent
as soon as it is correctly defined : thanks to the gauge
invariance of the action, we have proven that the normalisation
condition
$c^{\mu} = 0$ may be enforced to all orders of perturbation theory.

\item The 2-photon Green function receives
definite (as they are finite by power counting) radiative corrections \cite{Bonneau01} $$\quad \simeq
\frac{\hbar e^2}{12\pi^2}
\frac{p^2}{m^2} \epsilon_{\mu\nu\rho\si} \,p^{\si}b^{\rho} + \cdots
$$ Recall the case of the electric charge : physically measurable
quantities occur only through the
$p^2$ dependence of the photon self-energy (as the Lamb-shift is a
measurable consequence of a non-measurable charge renormalisation).
Unfortunately, as Coleman and Glashow explained, the absence of
birefringence of light in vacuum,
$i.e.$ the vanishing of the parameter $c^{\mu}\,,$ gives no
constraint on the value of the other one $b^{\mu}\,.$ However, in \cite{{Nasc},{Cervi}}, CPT breaking in QED is studied at finite temperature : as is clearly emphasized in \cite{Cervi}, the T dependance being independant on the normalisation condition at T= 0 , as soon as it is carefully computed it offers an unambiguous induced Chern-Simons like term at finite T and can give some information on $b^{\mu}\,.$
\end{itemize}

\noindent {\large {\bf Thanks :}} I have truly appreciated some
correspondance with M. P\'erez-Victoria.

\bibliographystyle{plain}

\end{document}